# Chiral exceptional point enhanced active tuning and nonreciprocity in micro-resonators

Hwaseob Lee[1†], Lorry Chang[1†], Ali Kecebas[2], Dun Mao[1], Yahui Xiao[1], Tiantian Li[1], Andrea Alù[3,4*], Sahin K. Özdemir[2,5*], Tingyi Gu[1*]

[1] Department of Electrical and Computer Engineering, University of Delaware, Newark, Delaware, 19716, USA

[2] Department of Engineering Science and Mechanics, Pennsylvania State University, University Park, PA 16802 USA

[3] Photonics Initiative, Advanced Science Research Center, City University of New York, New York, NY 10031, USA

[4] Physics Program, Graduate Center, City University of New York, New York, NY 10016, USA

[5] Department of Electrical and Computer Engineering, Saint Louis University, Saint Louis, MO 63103, USA

[†] Authors contributed equally

* Email: tingyigu@udel.edu, sko9@psu.edu, aalu@gc.cuny.edu

## Abstract

Exceptional points (EPs) have been extensively explored in mechanical, acoustic, plasmonic, and photonic systems. However, little is known about the role of EPs in tailoring the dynamic tunability of optical devices. A specific type of EPs known as chiral EPs has recently attracted much attention for controlling the flow of light and for building sensors with better responsivity. A recently demonstrated route to chiral EPs via lithographically defined symmetric Mie scatterers on the rim of resonators has not only provided the much-needed mechanical stability for studying chiral EPs, but also helped reduce losses originating from nanofabrication imperfections, facilitating the *in-situ* study of chiral EPs and their contribution to the dynamics and tunability of resonators. Here, we use asymmetric Mie scatterers to break the rotational symmetry of a microresonator, to demonstrate deterministic thermal tuning across a chiral EP, and to demonstrate EP-mediated chiral optical nonlinear response and efficient electro-optic tuning. Our results indicate asymmetric electro-optic modulation with up to 17dB contrast at GHz and CMOS-compatible voltage levels. Such wafer-scale nano-manufacturing of chiral electro-optic modulators and the chiral EP-tailored tunning may facilitate new micro-resonator functionalities in quantum information processing, electromagnetic wave control, and optical interconnects.

## Introduction

Non-Hermitian spectral singularities known as exceptional points (EPs), and the associated reduction in a system's dimensionality, have been extensively studied in mechanical, acoustic, plasmonic, and nanophotonic systems for their exotic response[1-3]. A specific type of EPs known as chiral EPs has been observed in ring resonators by properly positioning scatterers to perturb traveling wave resonator modes [4-7] or by terminating one end of a waveguide with a mirror[8-9]. Chirality here refers to the direction of rotation of the optical field inside the resonator. At a chiral EP the modal fields propagating in the clockwise (CW) or counterclockwise (CCW) direction become degenerate.[4] EPs can also emerge in parity-time (PT) symmetric systems by judiciously engineering the imaginary part of the refractive index of subsystems and their coupling. This corresponds to tuning the gain-loss balance of subsystems in active PT systems [10-14] and the loss-imbalance between subsystems in passive PT systems[15-18]. EPs have also been demonstrated in various quantum systems, including atomic ensembles[19], single spins[20], single trapped ions[21], and superconducting qubits[22]. Recently, electrostatic tuning of graphene permittivity has been implemented to achieve topological control of terahertz light across EPs[23-24]. The utility of EPs for achieving highly tunable systems, optical modulators, and enhanced light-matter interactions has not been studied thoroughly and it remains elusive. Elucidating this can potentially contribute to energy-efficient photonic precision instrumentations, such as analog processors[25], gyroscopes[26,] and atomic clocks.

The enhanced response of chiral EPs to small perturbations makes them appealing for sensing [27-28], efficient electro-optic signal transduction, optical interconnects, and isolators[29-30]. However, current implementations suffer from mechanical instabilities and fabrication-related imperfections. Overcoming these challenges and building mechanically stable chiral EP systems will allow precise control of critical parameters and enable the exploration of enhanced non-Hermitian photonic systems. To address these opportunities, we have fabricated a silicon photonic micro-ring resonator (MRR) with two lithographically defined asymmetric Mie scatterers (Fig. 1a). The two scatterers are geometrically engineered to introduce the same reflectance to the guided modes in one direction (Fig. 1b). By tuning only one of the two optical paths between the asymmetric scatterers with a highly localized heater, which is carefully aligned to one optical path along the resonator, we can control the intra-resonator coupling coefficient between the CW and CCW modes and thus move the system to an EP or away from it. As a result, we deterministically tune chirality, that is the direction of rotation of the optical field inside the resonator (Fig. 1c). We note that in this process, the lithographically defined scatterers redistribute the input optical power from transmission to reflection ports through coupling to the counter-propagating modes in the resonator, without introducing significant loss (Fig. 1c).

The realized non-Hermitian device thus can be deterministically steered between its EP degeneracy (i.e., the field inside the resonator is chiral and propagates in either the CW or the CCW direction and no mode-splitting in the spectra) and non-degenerate states (i.e., both CW and CCW propagating modes exist in the resonator and transmission spectra show mode-splitting). This dynamic control allows us to observe chiral EP which, together with EP-enhanced response of the system to small perturbations, improves phase-amplitude tuning sensitivity and leads to chiral nonlinear and modulation response (i.e., different responses for inputs in the CW and CCW directions). This is attributed to the fact that the asymmetric coupling between CW and CCW modes results in asymmetric field enhancement factors (EF), defined as the ratio of the inter-mode (CW-to-CCW or CCW-to-CW) coupling strengths to the total loss of the modes (Fig. 1d), which in turn leads to different nonlinear resonance shifts for CW and CCW inputs, and hence nonreciprocal response. The difference in the temporal dynamics and response of the system for CW and CCW inputs is clearly seen in Fig. 1e. In our device, we can achieve asymmetric electro-optic modulation at GHz speeds with 17dB contrast when the system is at the chiral EP, that is the light input in the CCW direction is modulated more than the light input in the CW direction.

Beyond electro-optic tuning and modulation, we also explored the EP's contribution to the chiral nonlinear response and nonreciprocities. The power range of such nonreciprocal switches is controlled by the chirality and intrinsic loss rate of the resonator (Fig. 1f), which can surpass the power range-transmission trade-off associated with single nonlinear resonators[31-32]. Low-energy and small-footprint silicon microring modulators are widely adopted for optical interconnects and neuromorphic computing[33-37]. If successful, this low-power and precision nanophotonic engineering can reduce the module redundancy in large-scale photonic integrated circuits for interconnects and computing.

**Results**

Non-Hermiticity is introduced into our waveguide-coupled microring resonator (MRR) system through asymmetric coupling between its frequency degenerate CW and CCW modes which is controlled by tuning the optical path length between two asymmetric Mie scatterers (i.e., inter-scatterer phase) via thermo-optic effect (Methods). The scatterers are lithographically defined within the resonator mode volume with dimensions between 1/6 and 1/3 times the effective wavelength in the single-mode silicon photonic waveguide. The geometric asymmetry of the scatterers is essential to realize chiral EP in a single MRR. Previously[38], we compensated asymmetric reflections of distributed Rayleigh scatterers (i.e., in the form of surface roughness or structural inhomogeneity formed during nanofabrication) by introducing a symmetric scatterer in the MRR and demonstrated back-reflection suppression and the emergence of an EP. Here we carefully designed the asymmetric scatterers to provide sufficient contrast between the reflection coefficients for CW and CCW propagating modes, while keeping the quality factors around $10^4$ (inset of Fig. 1a). The scatterer-induced coupling strength of the field into the same or the counterpropagating mode are described as different complex-valued elements ($\epsilon_2$ and $\epsilon_2'$, marked in the inset of Fig. 1a). Based on two-mode approximation and coupled mode theory, the effects of the two Mie scatterers and the inter-scatterer phase ($\Delta\varphi_v$) on the total Hamiltonian of the non-Hermitian MRR can be described as[39-40]:

$$H = \begin{pmatrix} \Omega_0 + \Delta\Omega_{\epsilon\_cw} + \Delta\omega_v & \epsilon^{Sym} + \epsilon^{Asym}_{ccw \to cw} e^{-j(\varphi_0 + \Delta\varphi_v)} \\ \epsilon^{Sym} + \epsilon^{Asym}_{cw \to ccw} e^{j(\varphi_0 + \Delta\varphi_v)} & \Omega_0 + \Delta\Omega_{\epsilon\_ccw} + \Delta\omega_v \end{pmatrix} = \begin{pmatrix} \chi_{11} + \Delta\omega_v & \chi^v_{12} \\ \chi^v_{21} & \chi_{22} + \Delta\omega_v \end{pmatrix} \quad (1)$$

Here $\Omega_0 = \omega_0 - i\gamma_t$ is the complex resonance frequency of the unperturbed (without the scatterers) MRR where $\gamma_t$ denotes the total loss, including the intrinsic (i.e., material, scattering, and radiation losses) and waveguide-resonator coupling losses, $\Delta\Omega_{\epsilon_{cw/ccw}} = \Delta\omega_{\epsilon_{cw/ccw}} - i\gamma_{\epsilon_{cw/ccw}}$ denote complex frequency change induced by the Mie scatterers in CW/CCW modes, and $\Delta\omega_{EO} = \Delta\varphi_v c/L$ is the frequency shift due to the local phase shift $\Delta\varphi_{EO}$ where $c$ is the velocity of light in the waveguide and $L$ is the perimeter length of the MRR. $\epsilon^{Sym}$ and $\epsilon^{Asym}$ in the off-diagonal elements of the Hamiltonian represent the complex coupling coefficients induced by the Mie-scatterers between CW and CCW modes and they are set by the geometry of the scatterers (Inset of Fig. 1a). The inter-mode coupling rates defined as $\chi^v_{12} = \epsilon^{Sym} + \epsilon^{Asym}_{ccw \to cw} e^{-j(\varphi_0 + \Delta\varphi_v)}$ and $\chi^v_{21} = \epsilon^{Sym} + \epsilon^{Asym}_{cw \to ccw} e^{j(\varphi_0 + \Delta\varphi_v)}$ can be tuned by varying the inter-scatterer phase $\varphi_0 + \Delta\varphi_v$ (Fig. 1b). Note that in Fig. 1, we have marked $\epsilon^{Sym}$, $\epsilon^{Asym}_{ccw \to cw}$, and $\epsilon^{Asym}_{cw \to ccw}$ as $\epsilon_1$, $\epsilon_2$ and $\epsilon_2'$, respectively. The eigenvalues of H in the absence of any external thermal effect are $\omega_\pm = \omega_0 + \Delta\omega_\epsilon - i\Gamma \pm \xi/2$ where $\xi = \sqrt{4\chi^v_{12}\chi^v_{21}}$ is the amount of mode-splitting. This expression assumes that the scatterer-induced frequency shifts are independent of input excitation, thereby preserving transmission reciprocity in the linear regime [36]. Moreover, it implies that by precisely tuning the local phase difference $\Delta\varphi_v$, we can steer the system to or from an EP ($\xi = 0$) (Supplementary Section 1). Other losses, such as ring-waveguide coupling losses ($\gamma_{c1/c2}$ in Fig. 1a) and radiation losses ($\gamma_{cw/ccw}$ in Fig. 1a) and thus $\gamma_t$, remain unchanged during the phase tuning.

Phase-only tuning affects the diagonal elements of $H$ in the same way as the term $\Delta\omega_{EO}$ whereas its effect on the off-diagonal elements differs significantly through the terms $e^{\pm j(\varphi_0+\Delta\varphi_v)}$. Moreover, it drifts the system away from the critical coupling condition and enhances the amplitude tuning efficiency (named optical modulation amplitude, or OMA)[41-42]. The enhanced OMA is evidenced by the large transmittance contrast between the initial and final states at the optimized detuning. The dependence of the resonance frequency shift and the amount of EP-enhanced mode-splitting on $\Delta\varphi_v$ suggests magnifying the phase-peak amplitude tuning by varying $\Delta\varphi$ (left red part in Fig. 1d). In this way, the amplitude modulation efficiency of the non-Hermitian MRR can be made to exceed that of regular MRR (no peak tuning) and of an MRR operating near the diabolic point (DP). In addition, the chiral response allows opposite tuning effects for light input in opposite directions (right grey part Fig. 1d): At a selected laser-cavity detuning, the electro-optic phase tuning results in enhanced amplitude response in one direction and minimized OMA in the other direction. Thus, the device functions as a chiral electro-optic modulator.

The design concept is numerically illustrated in Fig. 2a-c. The system can be switched from the non-EP state (mode-splitting with standing wave mode profile in Fig. 2a) to the EP state (traveling wave mode profile in Fig. 2b) when $\Delta\varphi_v$ is changed from $0.5\pi$ to $0.85\pi$ (marked in Fig. 2c). If the geometry of the scatterers ($\epsilon^{Sym}$ or $\epsilon^{Asym}$) is optimally selected using the optical impedance matching method, then one can continuously vary the inter-scatterer phase $\Delta\varphi_v$ to drive the system towards an EP (detailed inter-scatterer mode profile in the right inset of Fig. 2b), where both eigenmodes coalesce and feature a square root dependence on detuning ($|\Delta\omega|\sim|\sqrt{\epsilon}|$). We quantify whether the system is at the EP degeneracy or detuned from it using the standing wave ratio $\Gamma$ and the mode non-orthogonality parameter $S$ (definitions are given in equations S.2-1 and S.2-2, respectively). Near the EP degeneracy, we find $\Gamma\sim0.064$ and $S\sim1$ (derived from Fig. 2b) which imply that the two eigenvectors become collinear and the field inside the resonator is dominantly in one direction (i.e., standing wave ratio close to zero implies a traveling field) as the system approaches the EP. Experimentally, the inter-scatterer phase tuning is achieved by a local heater defined along the perimeter of the MRR. With calibrated electronic characteristics of the sub-µm width heater (Fig. 2d and Fig. S2a), the generated temperature profile for optical phase shift is confined within a few µm range between the scatterers defined on the perimeter of the MRR (Fig. 2e). Varying the inter-scatterer phase $\Delta\varphi_v$ tunes the eigenvalues (Fig. 2c) and the associated eigenvectors ($\psi_\pm$) (Fig. 2f and 2g). The eigenvectors collapse in the CW direction at the phase matching point (reference point of $\Delta\epsilon = 0$). For a scatterer geometry with $\Delta\epsilon \neq 0$, neither the eigenvalues nor eigenvectors cross each other when the inter-scatterer phase is varied (black curve in Fig. 2c and Fig. 2g).

We performed experiments and evaluated the performance of the phase-only tuning scheme for non-Hermiticity. The results of the experiments are given in Fig. 3. The center resonance wavelength of both CW and CCW excited modes linearly shifts with $\Delta\varphi_v$, but the off-diagonal elements move in opposite directions on the complex plane (Fig. 3a). By simultaneously fitting the coupled mode theory (CMT) to the experimentally measured transmission and reflection spectra (Supplementary Section 3), we extracted the voltage-dependent (i.e., voltage applied to the heater) $\chi_{12}^v$ and $\chi_{21}^v$ red and grey circles in Fig.3a). Initially, the system is at an EP, as $\chi_{12}^v$ locates at the origin and $\chi_{21}^v \neq 0$. As the voltage increases, $\chi_{12}^v$ rotates in the CCW direction, but $\chi_{21}^v$ rotates in the CW direction. On the transmission spectra, the mode-splitting increases with the amplitudes of the off-diagonal elements, and the non-zero phase of those off-diagonal elements lead to asymmetry in the deconvoluted modes (red and black dashed curves in Fig. 3b). For an exemplary mode centered near 1547 nm, we observe that the system, which is at the EP when the voltage applied to the heater is 0.1V (1.5mW of heating power), moves away from the EP as the applied voltage is gradually increased to 0.55V (local heating power around 4.5mW), resulting in an asymmetric mode-splitting. The dynamic tunability around EP (at low drive voltages of less than 0.2V) is evident by the high reflection contrast, weak reflection with CW excitation, and zero mode-splitting in the transmission spectra obtained for CW excitation (dark squares in Fig. 3c, d, e). The mode splitting ($\Delta\lambda$) is obtained by fitting the measured transmission spectra with a dual Lorentzian model. The full width at half maximum (FWHM) of the decomposed Lorentzian spectra for CW and CCW modes (red and black dashed curves in Fig. 3b) remains unchanged when the voltage applied to the heater is varied, indicating that the process does not change the loss rates ($\gamma_{CW/CCW}$) of CW and CCW modes. As a result, the quality factor remains ~$10^4$. In the same device under the same electrical drive, the other mode centered at 1542 nm (mode 2) evolves to the EP degeneracy when the voltage is increased from 0.25V to 0.55V (Fig. S2b-c, and grey dots in Fig. 3c-e).

We quantify the peak transmission contrast using (see derivation in Supplementary Section 4)

$$\frac{T_{EP}}{T_{non-EP}} = \left|1 + \frac{\chi_{12}^v \chi_{21}^v}{\gamma_t^2/4}\right|^2 \tag{2}$$

which suggests that the contrast can be maximized by precisely tuning the inter-mode coupling strength via $\Delta\varphi_v$ if the total loss does ($\gamma_t$) not vary significantly during the process. We observe that during the thermal tuning of $\Delta\varphi_v$, the waveguide-coupled resonator system moves from the under-coupling regime towards critical coupling at the EP degeneracy ($\Delta\lambda=0$), resulting in the highest transmission peak (red solid curve in Fig. 3f, at 0.35V). As the system deviates from the EP, we observe a significant reduction in peak transmission and an increase in the linewidth. We also observed that OMA near the EP (red curve in Fig. 3g) is twice as a non-Hermitian mode (inset of Fig. 3g), under the same initial and final drive voltages.

The presented low-loss chiral micro-ring resonator facilitates nonreciprocal signal routing in the nonlinear regime. The asymmetry in the field strengths for CW and CCW mode is manifested through asymmetric nonlinear cavity built-up (Fig. 4a-c). Fig. 4a-b shows the measured spectra of output power for increasing input power for CCW and CW excitations. The shift in the resonance wavelength for CW excitation ($\Delta\lambda_{cw}$) for increasing input power exceeds the one for CCW excitation ($\Delta\lambda_{ccw}$). A similar contrast of the nonlinear resonance shift is observed at the reflection ports (Fig. 4c) where we also see that optical bistability broadens the EP spectral range. To verify the nonreciprocal response, we compare the nonlinear transmission spectra $T_{1-2}$ (measure port 2 for input at port 1 for CW excitation), and $T_{2-1}$ (measure port 1 for input at port 2 for CCW excitation) (Fig. 4d). The nonlinearity-induced resonance shifts ($\Delta\lambda$) are proportional to the average intracavity field intensity or cavity energy [43-44] (Fig. 4d), so is the non-reciprocity intensity range (NRIR) given by:

$$NRIR = \frac{\gamma_t^2/4 + |\chi_{21}^v|^2}{\gamma_t^2/4 + |\chi_{12}^v|^2} \tag{3}$$

The nonlinear resonance shifts of the CW mode ($\Delta\lambda_{cw}$) exceeds the one for CCW excitation ($\Delta\lambda_{ccw}$) (Fig. 4f). Similar contrast of the nonlinear resonance shift is observed between ports 1 and 3 (Fig. 4e). Near EP ($\chi_{12}^v \cong 0$), the loss-limited maximum NRIR is found as $1 + \frac{|\chi_{21}^v|^2}{\gamma_t^2/4}$. The nonlinear chiral MRR is not subject to the loss-NRIR trade-off which limits the performance of typical single nonlinear Fano resonator systems (Fig. S3). Reducing the loss of the asymmetric Mie scatterer and the phase tuner is critical for facilitating a reasonable NRIR for nonreciprocal device performance. Figure 4g presents the theoretical predictions of NRIR versus chirality for various values of $\gamma_t/|\chi_{21}^v|$ together with the experimentally measured NRIR (tracked by comparing the reflection spectra) values. The NRIR can be further improved through finer tuning of the inter-scatter phase or reducing fabrication imperfections.

We have also demonstrated wafer-scale manufacturing of doped silicon photonic MRMs (Fig. 5a). The design concept, in particular the asymmetric Mie-scatterer and local phase shift, was first implemented with e-beam lithography (Supplementary Section 5), and then the chip was sent to standard semiconductor foundries to fabricate *p-n* junctions along the ridge waveguide between the scatterers for ultrafast local phase modulation (up to GHz). This combined with proper photonic engineering enabled us to observe chiral modulation controlled with an electronic drive, that is the optical carrier is modulated only in one direction (CW or CCW excitation) while the transmission from the other excitation direction remains unperturbed.

The directional MRM suppresses the side flow of data in an optical link, reduces the circuit design complexity and duplications, and supports recurrent routing topologies for interconnects and computing (Fig. 5b)[45-47]. When the excitation is in the CCW direction, the EP degeneracy in our system emerges at a lower voltage (0.5V) and the system moves away from the EP at a higher voltage (1V) resulting in mode-splitting in the transmission spectra. The bias-dependent chirality is the opposite for CW excitation (Fig. 5c). Time domain modulation confirms the unidirectional electro-optic modulator (Fig. 5d). The small transmission contrast ($T_{CW}/T_{CCW}$) measured under the steady state condition (Fig. 5c) is magnified under dynamic tuning ($\Delta T_{CW}/\Delta T_{CCW}$). Under the same RF electronic drive and DC bias, a vector network analyzer measured 17 dB contrast at modulation speeds up to GHz (Fig. 5e-f). The asymmetric high-speed OMA (optoelectronic S21) is verified at different detuning between laser and resonance wavelengths, showing the characteristic spectra of an OMA (Fig. 5f). The mode-splitting analysis of transmission spectra is detailed in Supplementary Section 6.

**Discussion**

Here we have demonstrated a chiral electro-optic modulator and nonreciprocal router in a silicon photonic platform by embedding low-loss asymmetric Mie scatterers in a microring resonator and by tuning the inter-scatterer optical path (i.e., phase) using a highly localized heater integrated to the resonator. Fine-tuning of the inter-scatterer phase steers the resonator controllably to and away from the EP and tunes the effective photon lifetime (i.e., the inverse of the total loss rate) of CW and CCW modes. We note that intrinsic effective loss rates for CW and CCW modes are given as $|\gamma_i|+|\chi_{12}^v|$ and $|\gamma_i|+|\chi_{21}^v|$, respectively, and the transmission is maximized at the critical point, where the intrinsic effective loss rate is equivalent to the coupling loss rate.

The EP enhancement on dynamic tunability and nonreciprocity is determined by the Mie scatterer-introduced inter-mode coupling strength versus total loss (from material absorption, radiation, and scattering). The critical dimension of around 50 nm for the tailored scatterers was achieved through a high resolution 193nm deep UV lithography manufacturing line, with optimized immersion lithography, projection mask optimization, and multi-layer photoresist coating design, followed by fine-tuned dry-etch (developed at AIM photonics). Different from the conventional EP systems, the radiation loss remains invariant with tuning, supported by the directly correlated reflection intensity and transmission mode-splitting. Tuning the optical path difference between the scatterers dynamically tunes the inter-scatterer phase which in turn controls the mode-splitting in different ways for CW and CCW modes. Simultaneous tuning of mode-splitting and resonance collectively enhances the electro-optic modulation response (quantified as OMA). Also, the dynamic tuning of the photon-loss channel expands the efficiency-bandwidth limit of the resonator-based modulator[48-49]. Higher $Q$ reduces drive voltage and energy, but the associated photon lifetime sets the upper limit of modulation speed. Dynamic tuning of $Q$ typically needs coupled resonators or coupled ring-waveguide schemes[50]. Here we adopted a conventional ridge waveguide phase modulator within a single ring, showing twice as high OMA at the EP compared to a reference mode of the same device that is far from the EP. Our method based on embedding asymmetric Mie scatterers can be extended to other types of modulators (e.g., Mach-Zender modulators) to introduce chirality. The chiral electro-optic tuning observed in our device can be further enhanced by more precise nanophotonic engineering and improved fabrication.

The scalability and applicability of the system are verified through a 300mm wafer-scale manufacturing, demonstrating GHz electro-optic bandwidth with 17dB contrast between the modulation amplitudes of CW and CCW excitations. The chiral response of the micro-ring circumvents the fundamental trade-off between insertion loss and the range of powers that support non-reciprocal transmission in Fano resonators, which relies on the feeding port-resonator asymmetric coupling. In our approach, the asymmetric scatter breaks the rotational symmetry of the resonator, and the chirality is manifested through nonlinearity and light-matter interactions for nonreciprocities beyond the loss-dynamic range trade-off in conventional nonlinear resonator isolators[51-52]. The interplay among chirality, non-Hermiticity, and resonance-enhanced nonlinear optical bistability introduces nonreciprocal optical signal routing in the fully passive silicon microring resonator. Such phase-sensitive electro-optic tuning, modulation, and all-optical nonreciprocity are facilitated through the strongly engaged CW and CCW modes. The additional degree of freedom on material absorption might need to be further analyzed by considering a multidimensional Hilbert space[53].

**Materials and methods**

**Nano-heater fabrication:** The MRR with local heater is fabricated on a silicon-on-insulator (SOI) from Soitec, having a 250 nm silicon layer and 3 μm buried oxide layer. On top of thick $SiO_2$ cladding, we fabricate a micro-heater (5 $\mu$m arc-length along the ring perimeter, and 50-920 nm width, with ~5nm alignment precision to the middle of the 420nm wide silicon wave under 700 nm oxide cladding). The metal is evaporated (5 nm Ti adhesion layer and 100 nm Pt heater) and the heaters are patterned by electro-beam lithography and double resist lift-off process.

**Foundry-manufactured chiral silicon photonic modulator and measurement:** The chiral MRMs were manufactured by AIM photonics through a multi-project wafer run (220 nm SOI). The lateral *p-n* diode configurations were defined by ion implantations: boron for *p*-type and phosphorus for *n*-type. Heavily doped *p++* and *n++* regions were used to form Ohmic contact, which connected the doped region through *p+* and *n+* region. Vertical *vias* are patterned and etched on cladding oxide for the contact regions, followed by standard aluminum metallization for direct contact with the heavily doped Si regions. The photonic structures were defined by 193 nm deep-ultraviolet photolithography on an 8-inch SOI wafer with a 220 nm device layer, followed by reactive ion etching. Three-step etching leaves silicon wing area for supporting the doping and contacts. A thick oxide cover layer is deposited for metal insulation. The RF-photonic measurements are detailed in a recent work[54].

**Funding:** This work was supported by the Defense Advanced Research Projects Agency (N660012114034). H.L. acknowledges the scholarship provided by the Republic of Korea Navy (ROK Navy). S.K.Ö. acknowledges the Air Force Office of Scientific Research (AFOSR) Multi-University Research Initiative (FA9550-21-1-0202) and AFOSR (FA9550-18-1-0235). A.A. acknowledges the Vannevar Bush Faculty Fellowship, the Air Force Office of Scientific Research MURI program and the Simons Foundation. The design and fabrication of the micro-heater and chiral MRR are supported by AFOSR (FA9550-18-1-0300).

**Author contributions:** H.L. and T. L. optimized the fabrication recipe and fabricated the samples. H. L., A. K., L. C., S.K.O., A.A. and L. S. developed the theoretical models. D.M. and T.K. built the experimental setup. L. C. and D. M. designed the layout for foundry manufacturing. H.L. and L. C. performed the measurements and analyzed the data, advised by A. A., S. K.O. and T. G. H.L., L.C., A. A., S.K.O., and T.G. prepared the manuscript, with contributions from all the authors.

**Acknowledgments:**

The authors acknowledge the inspiring discussions with Dr. Y. K. Chen for the chiral modulator and Dr. C. Hao from the Georgia Institute of Technology for the applications in optical interconnects. The devices with local heaters are fabricated at the University of Delaware Nanofabrication Facility.

**Competing interests**

The authors declare no competing interest.

**Data availability**

The data are available upon request.

# Figures

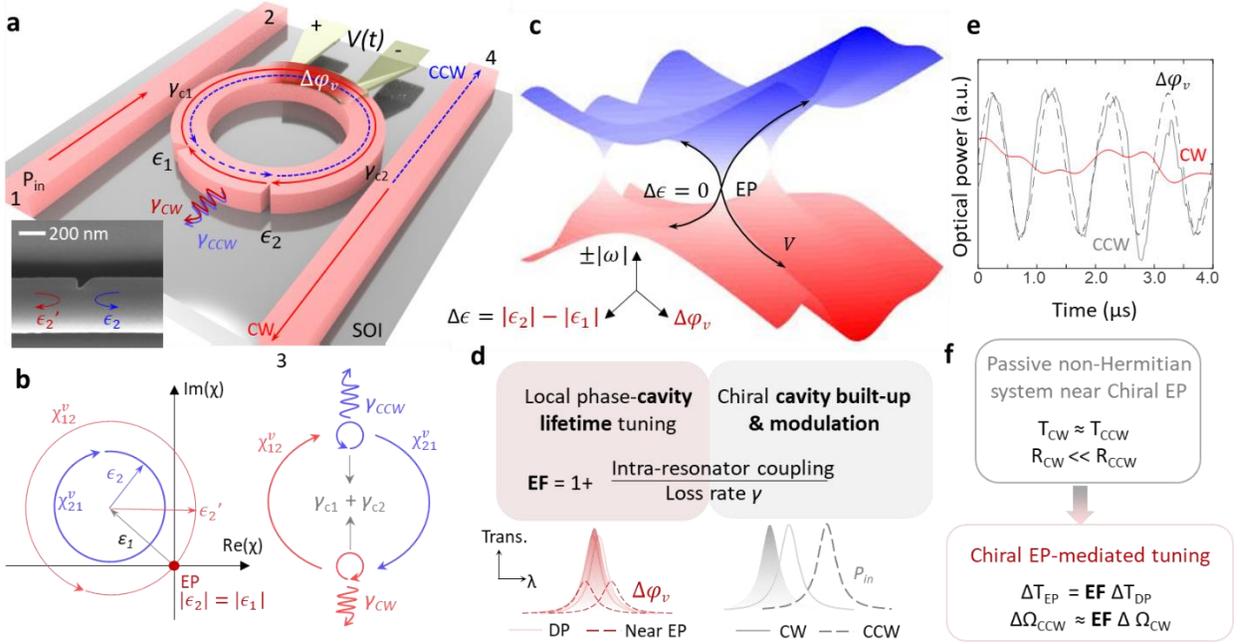

**Fig. 1 | Broken rotational symmetry mediated active tuning. a.** Schematics of the chiral microring resonator (MRR), with lithographically defined low loss asymmetric Mie scatterer (Inset), which mediates asymmetric coupling between clockwise (CW) and counterclockwise (CCW) modes. **b.** Asymmetric coupling strength ($|\chi_{12}^v|$ and $|\chi_{21}^v|$) tuned by the phase ($\Delta\varphi_v$) determined by the optical path between Mie-scatterers. An exceptional point (EP) emerges when the perturbation strengths of Mie scatterers are matched ($\Delta\epsilon = |\epsilon_1| - |\epsilon_2| = 0$). The complex inter-mode coupling contributes to the photon loss rate of the CW or CCW modes and is tunable through the inter-scatterer phase. The radiation loss rates ($\gamma_{cw/ccw}$) and coupling rate to waveguides ($\gamma_c$) are pre-set by the waveguide geometry (grey). For simplicity, $\epsilon^{Sym}$ in equation (1) is marked as $\epsilon_1$. $\epsilon_2$ and $\epsilon_2'$ are correspond to $\epsilon_{ccw \to cw}^{Asym}$ and $\epsilon_{cw \to ccw}^{Asym}$, respectively. **c.** Mode splitting versus perturbation strength offset ($\Delta\epsilon$) between the Mie scatterer pair and inter-scatter phase ($\Delta\varphi_v$). **d.** Enhancement factor (EF) from the chiral EP maximized with the coupling rate to loss rate ratio. Compared to the resonance tuning in diabolic point (DP), the highly localized heater efficiently tunes the coupling strength and cavity lifetime, and thus the peak transmission. The upper bound of chiral contrast between CW and CCW excitations is set by EF. **e.** Experimentally measured chiral response with the same time-varying phase, with EF up to 17dB. **f.** The tunable chiral EP facilitated investigations (red), compared to studies allowed in passive systems (unstable or not tunable).

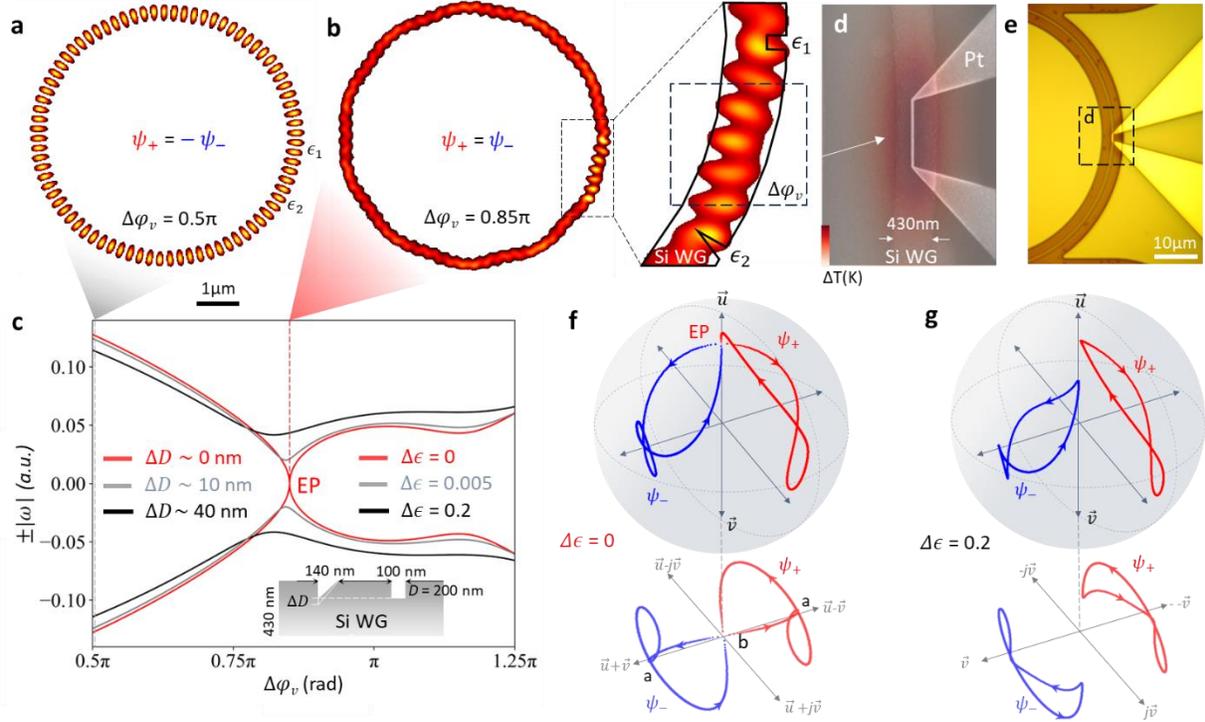

**Fig. 2 | Achieving EP with matching perturbation strength between a pair of Mie scatterers and deterministic tuning across EP with a nanoscale local heater.** Asymmetric Mie scatterers (Inset of Fig. 1d) pinned mode profiles of a micro-ring, at **a.** a non-EP state ($\Delta\varphi_v = 0.5\pi$, $\Gamma \sim 1$, $S \sim 0.06$) and **b.** an EP state ($\Delta\varphi_v = 0.85\pi$, $\Gamma \sim 0.064$, $S \sim 1$). Right inset: A closer view of the mode profile super-imposed on the Mie scatterers that supports EP. **c.** Mode splitting versus inter-scatter phase, with the Mie-scatterers geometry achieving EP (red, $\Delta\epsilon = 0$), Mie-scatterer perturbation offset of $\Delta\epsilon = 0.005$ (correspondent to the notch depth offset $\Delta D \sim 10$ nm) and $\Delta\epsilon = 0.2$ ($\Delta D \sim 40$ nm). Inset: Illustration of geometric offset $\Delta D$ given the scatterer and waveguide widths, for the 250nm SOI. **d.** Scanning electronic microscope image of the local heater superimposed with the resulting temperature distribution, illustrating the localized thermo-optic phase shift. **e.** Optical microscope image of a microring with a local heater. **f.** Representation of the evolution of the eigenvectors on the Bloch sphere by tuning inter-scatterers phase $\varphi_v$ only. The bottom insects are their projected trajectories for improving clarity. The Mie scatterers combination supports EP ($\Delta\epsilon = 0$), and **g.** does not support EP ($\Delta\epsilon = 0.2$). $\vec{u}$: eigenvector for the CW mode, and $\vec{v}$: CCW mode.

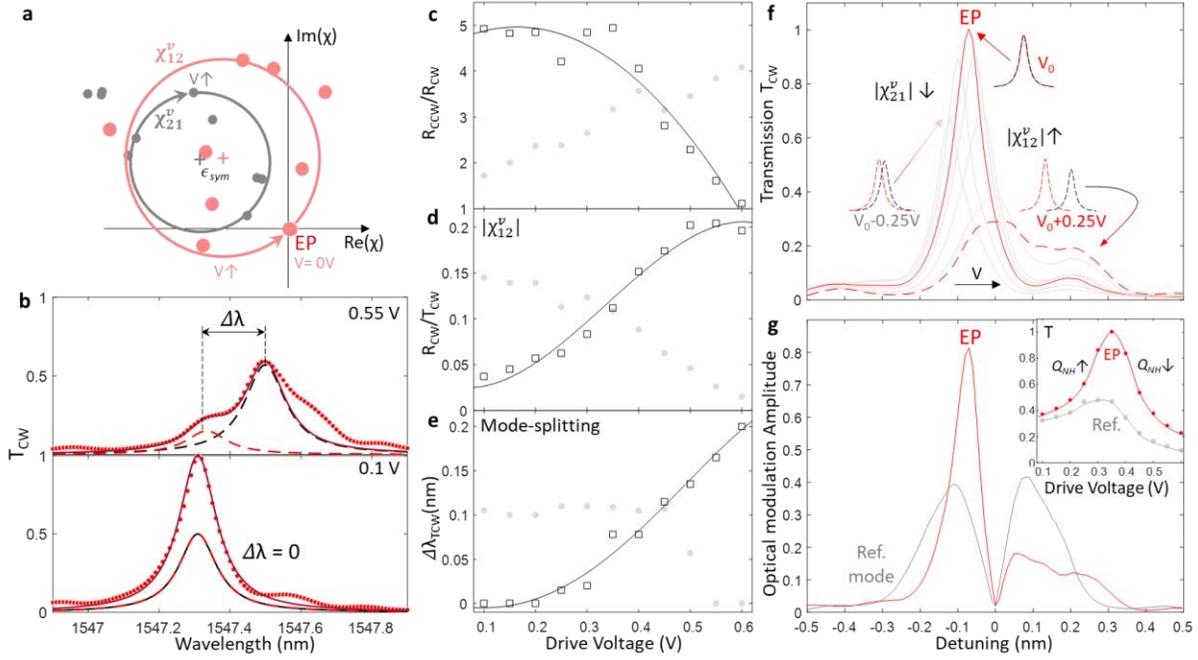

**Fig. 3 | Deterministic tuning of inter-scatterer local phase across EP. a.** Evolution of the off-diagonal elements ($\chi_{21}^v$ and $\chi_{21}^v$) as a function of the inter-scatter phase. The centers of the circles represent the complex coupling coefficient $\epsilon^{Sym}$. The center's offset is attributed to the surface roughness. The radius is $|\epsilon_{ccw \rightarrow cw}^{Asym}|$ for $\chi_{21}^v$ circle, and $|\epsilon_{cw \rightarrow ccw}^{Asym}|$ for $\chi_{21}^v$ circle. EP is achieved at $|\chi_{12}^v| = 0$, as $\chi_{21}^v$ never crosses the origin. **b,** Transmission spectra of a mode at two different voltages applied to the heater to tune the inter-scatter phase. The red dots are experimental data, which are analyzed by the coupled mode theory and empirical model of double Lorentzian fitting for deconvoluting the two dressed states. The additional side peak (near 1547.6 nm) might be from the reflections from bus waveguide terminals. **c,** Reflection intensity contrast for CW and CCW excitations. **d,** Normalized reflection strength to transmission. **e,** Mode-splitting observed in the transmission spectra (in panels **a** and **b**) as a function of the drive voltage. Empty squares: experimental data of the target mode in **a**. Grey solid dots: an alternative mode of the same MRR for comparison. Curves are eye guides. **f,** Electro-optic tuning enhancement through coordinated diagonal and off-diagonal tuning. **g,** Contrast of transmission $|T(0.35V)/T(0V)|$ of a chiral mode (red) and a reference mode (grey) for the same resonator. Inset: Transmission versus drive voltage for the chiral mode (red) and the reference mode (grey).

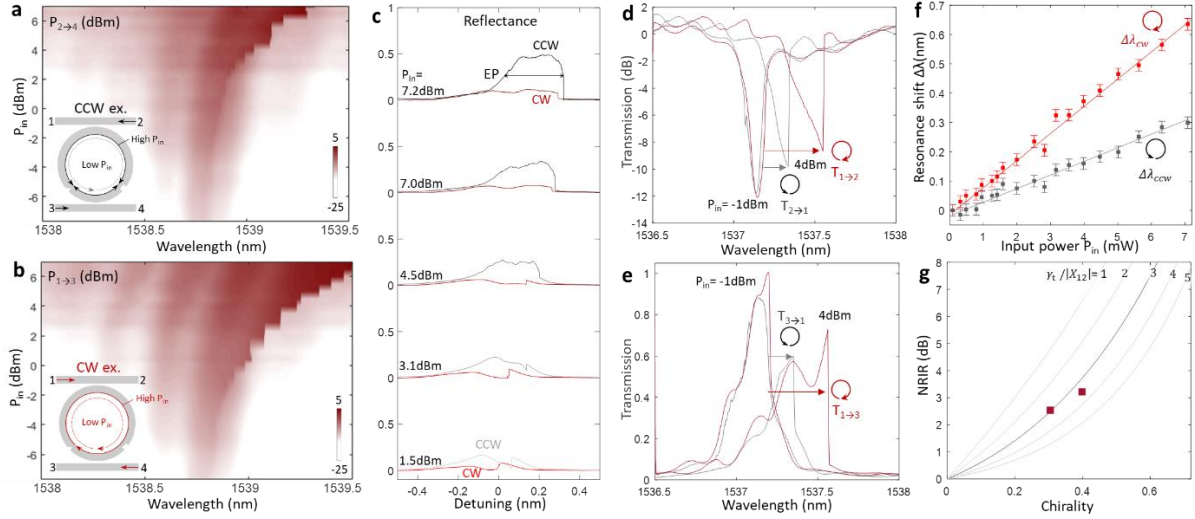

**Fig. 4 | Nonlinearity manifested chirality and nonreciprocity**. **a.** Input power-dependent transmitted output spectra, with CCW excitation, and **b.** CW excitation, at increasing input optical power ($P_{in}$). Insets: mode evolution at low (dashed) and high excitation powers (solid), for CCW (in panel a) and CW (in panel b) excitations. **c.** Correspondent reflection spectra for CW (red) and CCW (grey) excitations, show expanded EP bandwidth through optical bistability. **d.** Nonreciprocal transmission between through ports (1 and 2) **e.** Transmission between drop ports 1 and 3 with CW (red) and CCW (grey) excitations. Their transmission spectra nearly overlap at low power ($P_{in}$ = -1dBm) but have divergent optical nonlinear responses ($P_{in}$ = 4dBm). **f,** Nonlinear optical resonance shifts versus input optical power with CW (red) and CCW (grey) excitations. **g,** The dependence of NRIR on the MRR chirality.

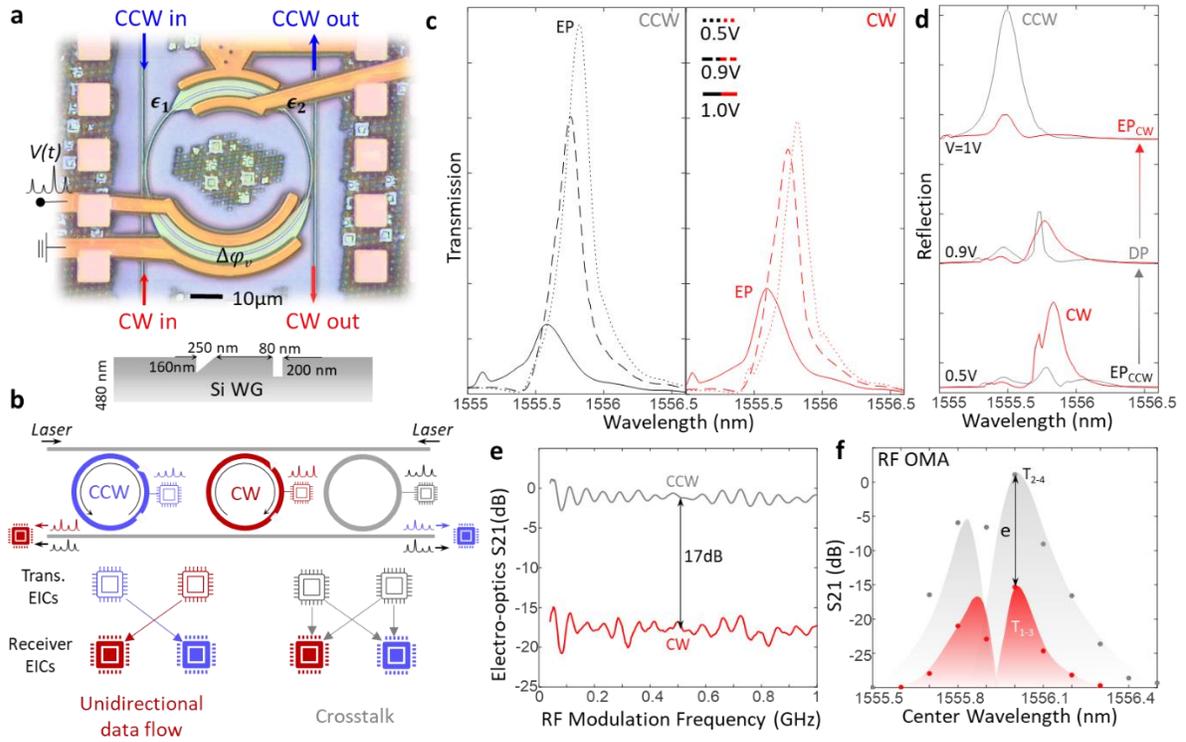

**Fig. 5 | EP enhanced GHz chiral electro-optic modulator. a,** 3D imaging laser microscope captured image of the non-Hermitian photonic modulators coupled to add/drop bus waveguides embedded in oxide, manufactured through a foundry wafer run. Inset: scatterer designs for the 220nm SOI. **b,** Exemplary application scenario of the chiral electro-optic response enabled unidirectional data flow between transmitting electronic integrated circuits (EICs) and receiving EICs. In conventional MRMs, duplicated circuit topologies for transmitting and receiving terminals are required to avoid crosstalk (grey). **c.** Measured transmission spectra at CCW excitation (left) and CW (right) excitation with increasing electrical bias voltage across the *p-n* junctions of GHz phase modulators. The free-carrier absorption modulation is superimposed on the non-Hermiticity tuning response (illustrated in **b**). **d.** Correspondent reflection spectra at increasing bias voltage, implying the chirality of the device. **e.** Electro-optic S21 (normalized OMA measured at high speed) versus electrical modulation speed for CW (red) and CCW (grey) excitations. **f.** S21 versus wavelength for CW and CCW excitations.
191919

# Supplementary Material for

**Chiral exceptional point enhanced active tuning and nonreciprocity in micro-resonators**

Hwaseob Lee[1†], Lorry Chang[1†], Ali Kecebas, Dun Mao, Yahui Xiao, Tiantian Li, Andrea Alù[*], Sahin K. Özdemir[*], Tingyi Gu[*]

[†] Authors contributed equally

Email: tingyigu@udel.edu, sko9@psu.edu, aalu@gc.cuny.edu

Supplementary Text

Sec 1: Design of Mie scatterers for unidirectional coupling

Sec 2: Evolution of the mode profiles across an exceptional point

Sec 3: Inter-scatterer phase-controlled mode splitting

Sec 4: Chiral EP enhanced active tuning

Sec 5: Nanofabrication and characterizations

Sec 6: Nanomanufacturing chiral micro-resonator

Figures S1 to S12

Table S1

**Supplementary Section 1: Design of Mie scatterers for low loss & unidirectional coupling**

The dimensions of Mie scatterers were selected to provide sufficient coupling strength, without reducing the radiation-limited quality factor of the resonator. Performing three-dimensional full-field simulation and optical impedance matching (detailed in Section S1 of the reference [S1]), we located the embedded Mie scatterers such that exceptional points (EPs) emerge. Full field simulation of microring resonator (MRR) with embedded scatterers design confirms that the transmission spectra obtained for clockwise (CW) and counterclockwise (CCW) excitations are different. The conformal mesh with a spatial resolution (in $\hat{x}$ and $\hat{y}$ direction) less than 1/10 of the Mie scatterer dimension is applied. For the vertical direction ($\hat{z}$), the spatial resolution of the mesh is fixed as 1/5 of the simulated silicon waveguide thickness. The simulated waveguide structure is extended completely through the perfectly matched layer for stable and accurate results. Here we have numerically implemented two types of Mie scatterer combinations achieving EP (Fig. S1a-b). The exemplary symmetric and asymmetric Mie scatterers are illustrated in Fig. S1c-d.

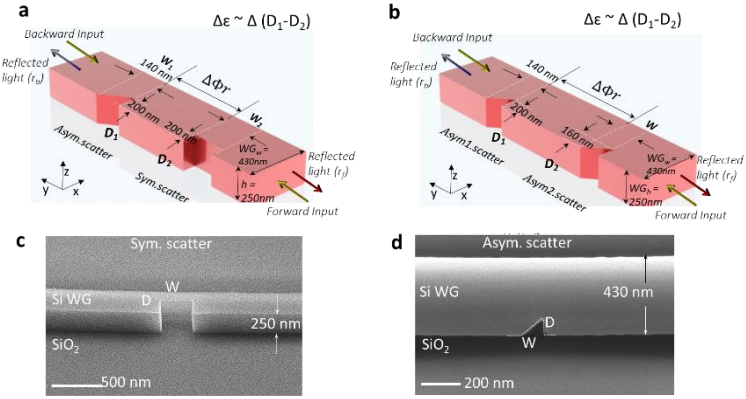

**Fig. S1 | Ideal design of the Mie scatterer pair supporting EP (without surface roughness). a.** Symmetric and asymmetric. **b.** Two asymmetric ones. The offset between two Mie scatterers is proportional to Δε in Fig. 1. **c.** SEM image of exemplary symmetric scatterer and **d.** asymmetric scatters.

The coupling strength of the $i$-th Mie scatter-induced scattering into the same ($k = m$) or the counterpropagating ($k \neq m$) mode is described as complex-valued elements $\epsilon_{ikm}$ (Fig. S1). The Hamiltonian describing the dynamics of a MRR with two embedded Mie Scatterers as function of the strength of perturbation $\epsilon_{ikm}$ induced by the Mie scatterers and the relative phase $\Delta\varphi$ (determined the optical path difference between the scatterers) between them is given by

$$H_2 = \begin{pmatrix} \epsilon_{111} + \epsilon_{211} & \epsilon_{112} + \epsilon_{212} e^{-j\Delta\varphi v} \\ \epsilon_{121} + \epsilon_{221} e^{j\Delta\varphi} & \epsilon_{122} + \epsilon_{222} \end{pmatrix} = \begin{pmatrix} \Delta\omega_o + \chi_1 & \chi_{12} \\ \chi_{21} & \Delta\omega_o + \chi_2 \end{pmatrix} \quad \textbf{(S. 1-1)}$$

The diagonal elements $\epsilon_{i11}$ and $\epsilon_{i22}$ determine the resonance frequency shifts induced by the $i$-th scatters in the CW and CCW modes. The low loss Mie scatter geometries ensure nearly vanished imaginary parts for $\epsilon_{i11}$ and $\epsilon_{i22}$, and thus they do not induce additional losses into the CW and CCW modes. The off-diagonal complex elements $\epsilon_{i12}$ and $\epsilon_{i21}$ determine the i-th scatter induced coupling between CW and CCW modes. The eigenvalues of $H_2$ are $\omega_\pm = \omega_0 + (\chi_1 + \chi_2)/2 \pm \xi/2$ where $\xi = \sqrt{(\chi_1 - \chi_2)^2 + 4\chi_{12}^v \chi_{21}^v}$. Clearly, $\xi = 0$ results in coalescing eigenvalues (EPs). Corresponding eigenvectors are also computed as: $\psi_\pm = [(\chi_1 - \chi_2 \mp \xi)/2\chi_{21} \quad 1]^T$. When $\xi = 0$, both eigenvectors and corresponding eigenvalues coalesce simultaneously, indicating emergence of an EP. In order to achieve EP condition, first, from equation (1), $\xi$ is explicitly expressed in terms of $\epsilon_{ikm}$ and $\Delta\varphi$ as: $\xi^2 = (\chi_1 - \chi_2)^2 + 4[\epsilon_{121}\epsilon_{112} + \epsilon_{212}\epsilon_{221} + (\epsilon_{121}\epsilon_{212} + \epsilon_{112}\epsilon_{221})\cos\Delta\varphi + j(\epsilon_{112}\epsilon_{221} - \epsilon_{121}\epsilon_{212})\sin\Delta\varphi]$. For an EP to emerge at $\xi = 0$, the first step is selecting the geometry of the scatterers such that $\epsilon_{112}\epsilon_{221} = \epsilon_{121}\epsilon_{212} = \epsilon'$. With fixed $\text{Im}(\xi^2) = 0$, the mode splitting varies periodically with the electrically tunable $\Delta\phi$. EPs emerge at $\text{Re}(\xi^2) = 0$, where $\cos\Delta\varphi = -\frac{1}{2}\left(\frac{(\chi_1-\chi_2)^2}{4\epsilon'} + \frac{\epsilon_{112}^2 + \epsilon_{212}^2}{\epsilon_{112}\epsilon_{212}}\right)$. Therefore, one can conclude that it is possible to realize an EP through tuning the shapes and dimensions of the scatterers and the relative position between them.

**Supplementary Section 2: Evolution of the mode profiles across an exceptional point**

With a TE mode injected to the waveguide, the magnetic field is perpendicular to the cavity plane and the field can be expanded in cylindrical harmonics (Fig. S2a). The field data in numerical simulation are acquired by having separate spatial distances (*L*) of 1.4715μm and 1.5915μm between two scatters, corresponding to the optical phase delay $0.85\pi$ and $0.5\pi$, computed from: $\Delta\varphi = \frac{2\pi}{\lambda_{eff}} \times L$. This explains that the field can be varied periodically by either physically changing the spatial distance between scatters through an elaborate nanofabrication process or precisely controlling the phase delay by tuning silicon's refractive index (Fig. 2). The field data in the numerical simulation assesses the circularly distributed magnetic field data in the ring resonator when *R = R_o*, shown in Fig. S2b-d. Fig. S2d illustrates the average field intensity along the circular path, where near the degeneracy ($\Delta\varphi v = 0.85\pi$), the intensity is twice stronger than the intensity extracted from standing mode ($\Delta\varphi v = 0.5\pi$).

Moreover, the mode non-orthogonality and standing wave ratio are computed based on the extracted field data. This computation serves to substantiate the attainability of an exceptional point within our engineered scattering system. The nonorthogonality is quantified by [S2, S3]:

$$S = \frac{|\int dxdy \psi_1^* \psi_2|}{\sqrt{\int dxdy \psi_1^* \psi_1} \sqrt{\int dxdy \psi_2^* \psi_2}} \qquad \textbf{(S. 2-1)}$$

where $\Psi_{1(2)}$ is the extracted field data corresponding two eigenmodes of split modes $\omega_{+(-)}$

respectively. S is zero for orthogonal states and it is one for collinear states (e.g., at the EP). The calculated non-orthogonality corresponding to $\Delta\varphi = 0.85\pi$ and $\Delta\varphi = 0.5\pi$ are 1 and 0.401 respectively, which implies that eigenvectors of the system become collinear when $\Delta\varphi$ is properly controlled to steer the system to the vicinity of an EP.

In addition, we find the standing wave ratio [S4]

$$\Gamma = \frac{\sqrt{I_{max}} - \sqrt{I_{min}}}{\sqrt{I_{max}} + \sqrt{I_{min}}} \quad \text{(S. 2-2)}$$

as 0.06 and 1 for $\Delta\varphi = 0.85\pi$ and $\Delta\varphi = 0.5\pi$ respectively. In conclusion, this investigation affirms that our meticulously engineered MRR with two Mie scatters can achieve the degeneracy through a judiciously adjusted spatial separation between two scatters or the precise control of relative phase delay between them. As the system approaches the EP degeneracy, the eigenmodes of the system become non-orthogonal, finally coalescing (becoming collinear) at the EP. Similarly, as the system approaches the EP, the standing wave ratio decreases, becoming zero at the EP (i.e., traveling wave at the EP), This in turn provides a chiral wave propagation in the resonator and leads to maximal field intensity. These numerical results further confirm that the degeneracy obtained by adjusting the optical phase delay between two scatters is indeed an EP [S2, 3].

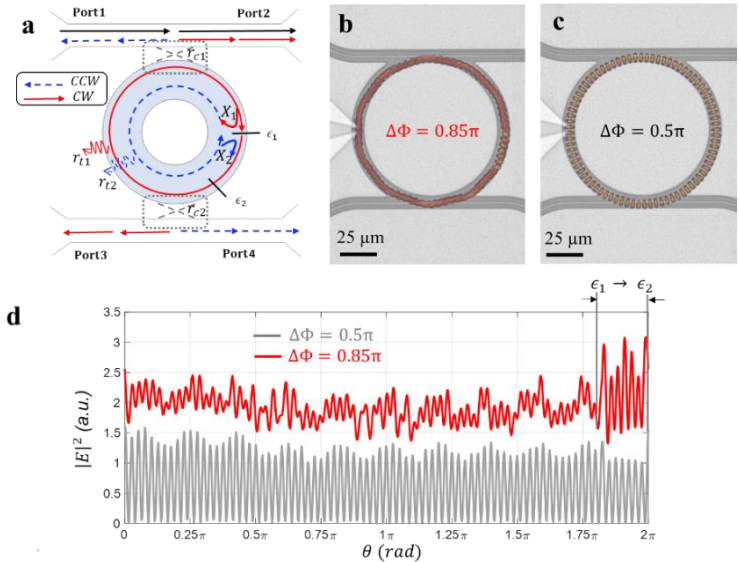

**Fig. S2 | Analytical model and numerical simulation.** (a) Schematic diagram of two scatterers perturbing the intracavity field of the MRR configured as an add-drop filter. The perturbation strengths of the asymmetric scatters are represented as $\epsilon_1$ and $\epsilon_2$. (b) Traveling mode-like profile with an optimized optical path delay of $\Delta\varphi = 0.85\pi$ provided by a local micro-heater. (c) Standing mode profile with an optical path delay of $\Delta\varphi = 0.5\pi$ provided by a local micro-heater. (d) Mode distribution along the circular direction of ring ($\theta$), for which each field data is extracted alone the center of the waveguide.

**Supplementary Section 3: Inter-scatterer phase-controlled mode splitting**

In general, the micro-heater is introduced in silicon MRR to shift the resonance in different applications such as optical switching and wavelength division multiplexing. Here, a micro-heater is adapted to induce a change in the real part of silicon refractive index through thermo-optic effect and demonstrate the enhancement of resonance shift in Fig. 3 of main text. Here, we discuss the enhancement of resonance shift. The initial (zero bias) Hamiltonian $H_0$ of a MRR composed of one symmetric and one asymmetric scatterer can be expressed as:

$$H_0 = \begin{pmatrix} \Omega_o + \Delta\omega_\epsilon & \epsilon^{Sym} + \epsilon^{Asym}_{ccw \to cw} e^{-j\Delta\varphi} \\ \epsilon^{Sym} + \epsilon^{Asym}_{cw \to ccw} e^{j\Delta\varphi} & \Omega_o + \Delta\omega_\epsilon \end{pmatrix} \quad \text{(S. 3-1)}$$

The two scatterers induce a total resonance shift of $\Delta\omega_\epsilon$ in both CW and CCW modes as well as create asymmetric coupling between the modes. After adding the electro-optic tuning terms, the updated Hamiltonian ($H$) can be expressed as.

$$H = \begin{pmatrix} \Omega_o + \Delta\omega_\epsilon & \epsilon^{Sym} + \epsilon^{Asym}_{ccw \to cw} e^{-j\Delta\varphi} \\ \epsilon^{Sym} + \epsilon^{Asym}_{cw \to ccw} e^{j\Delta\varphi} & \Omega_o + \Delta\omega_\epsilon \end{pmatrix} + \begin{pmatrix} \Delta\omega_{EO} & \epsilon^{Asym}_{ccw \to cw} e^{-j\Delta\phi_{EO}} \\ \epsilon^{Asym}_{cw \to ccw} e^{j\Delta\phi_{EO}} & \Delta\omega_{EO} \end{pmatrix}$$

$$= \begin{pmatrix} \Omega_o + \chi & \chi_{12} \\ \chi_{21} & \Omega_o + \chi \end{pmatrix} \quad \text{(S. 3-2)}$$

where the diagonal elements $\Omega_0 = \omega_o - \frac{j\gamma_t}{2}$ represent the complex energies of the degenerate CW and CCW modes, $\gamma_t$ representing all losses including both the intrinsic losses ($\gamma_0$, including material, bending, ring geometry dependent loss) and the resonator-waveguide coupling losses rates ($\gamma_c$). $\Delta\omega_{EO}$ is the resonance shift induced by the external electro-optic tuning. The total resonance shift while a microheater in operation comprises of the absolute value of complex frequency splitting and original resonance shift, which is equivalent to the regular diabolic point MRR ($\Delta\lambda_o = \Delta\lambda_{DP} = \frac{2\pi c}{\Delta\omega_{EO}}$), thus $\Delta\lambda_{tot} = 0.5|\Delta\lambda_{eigen}| + \Delta\lambda_o$, where $\left|\frac{\Delta\lambda_{eigen}}{2}\right|$ is derived from equation (S.3-1):

$$\left|\frac{\Delta\lambda_{eigen}}{2}\right| = \frac{\lambda_0}{\omega_0}|\sqrt{\chi_{21}\chi_{12}}| = \frac{\lambda_o}{\omega_o}\left|\sqrt{\left(\epsilon^{Sym} + \epsilon^{Asym}_{cw \to ccw} e^{j\Delta\varphi}\right)\left(\epsilon^{Sym} + \epsilon^{Asym}_{ccw \to cw} e^{-j\Delta\varphi}\right)}\right| \quad \text{(S. 3-3)}$$

Assuming that the intrinsic backscattering perturbation is much greater than the perturbation strength ($\chi_{12} \gg \chi_{21} \sim 0$), the equation shows the resonance shift enhancement is maximized when the system reaches EP ($\chi_{21}=0$).

**Supplementary note 4: Chiral EP enhanced active tuning**

**4.1 Time domain coupled mode theory**

Within this section, we will explore the phenomenon of enhanced nonlinearities by leveraging the asymmetry in optical power distribution, supporting the results presented in Fig. 4. Intracavity field intensity-dependent nonlinearity is characterized mainly by the resonance shift, followed by the extinction ratio and line-shape of the transmission spectrum.

We extend the schematics in Fig. S2a with nonlinear effects [S1]. The general coupled-mode theory that governs the two counter-propagating modes (CW and CCW) are:

$$\frac{da_{cw}}{dt} = j(\Delta\omega)a_{cw} - \frac{\gamma_t}{2}a_{cw} - j\chi_{12}^v a_{ccw} - \gamma_{c1}\sqrt{P_{CW}} \quad \textbf{(S. 4-1)}$$

$$\frac{da_{ccw}}{dt} = j(\Delta\omega)a_{ccw} - \frac{\gamma_t}{2}a_{ccw} - j\chi_{21}^v a_{cw} - \gamma_{c2}\sqrt{P_{CCW}} \quad \textbf{(S. 4-2)}$$

where $a_{cw/ccw}$ is the amplitude of the CW and CCW propagating mode; $\kappa$ is the coupling coefficient between waveguide and cavity, adjusted by the background Fabry-Perot (FP) resonance in the waveguide; $P_{in,cw(ccw)}$ is the incident power; $\gamma_{tot}$ is the total loss rate that CW mode and CCW mode experience, which are equal when nonlinear effects are ignored; $X_{12(21)}$ is the complex reflection coefficient from the scatters in the resonator, resulting in the modal coupling from CW (CCW) mode to CCW (CW) mode; and $a_{in,\ cw(ccw)}$ are the input optical power injected from Port 1(2) in Fig. S2a. By substituting $a_{cw(ccw)} = A_{cw(ccw)}e^{-j\omega t}$ and considering steady state ($\frac{d}{dt}A_{cw(ccw)} = 0$), where $\Delta\omega = \omega_L - \omega_c$ is the detuning between the laser frequency ($\omega_L$) and cold cavity resonance ($\omega_c$).

The transmission spectra of the resonator for CW and CCW excitations can be derived as the following [S1]:

$$T_{cw} = \left|\frac{s_3}{s_1}\right|^2 = \left|\frac{\sqrt{\gamma_{c1}\gamma_{c2}}}{\frac{\chi_{12}^v \chi_{21}^v}{i\Delta\omega - \gamma_t/2} + i\Delta\omega - \gamma_t/2}\right|^2 \quad \textbf{(S. 4-3)}$$

$$T_{ccw} = \left|\frac{s_4}{s_2}\right|^2 = \left|\frac{\sqrt{\gamma_{c1}\gamma_{c2}}}{\frac{\chi_{12}^v \chi_{21}^v}{i\Delta\omega - \gamma_t/2} + i\Delta\omega - \gamma_t/2}\right|^2 \quad \textbf{(S. 4-4)}$$

As a passive device, the transmission spectra for CW and CCW excitations are nearly the same. Here we explicit the magnified nonreciprocal tuning and nonlinear response as the following.

**4.2 Enhanced local phase – amplitude tuning**

Near the exceptional point, one of the off-diagonal elements is zero. The transmission spectrum is written as $T = \left|\frac{\sqrt{\gamma_{c1}\gamma_{c2}}}{i\Delta\omega - \gamma_t/2}\right|^2$, with a peak transmission of $T_0 = \left|\frac{\sqrt{\gamma_{c1}\gamma_{c2}}}{\gamma_t/2}\right|^2$. The local phase tuning deviates the system from EP, reducing the peak transmission of $T_v = \left|\frac{\sqrt{\gamma_{c1}\gamma_{c2}}}{\frac{\chi_{12}^v \chi_{21}^v}{\gamma_t/2} + \gamma_t/2}\right|^2$. The peak transmission contrast (between regular MRM and EP MRM) near EP:

$$\frac{T_0}{T_v} = \left|1 + \frac{\chi_{12}^v \chi_{21}^v}{\gamma_t^2/4}\right|^2 \quad \textbf{(S. 4-5)}$$

Furthermore, from Eq. (S.3-3) it is evident that the relative phase between the scatterers, $\Delta\varphi$, which can be controlled by applying voltage bias, has significant effect on $\Delta\lambda_{eigen}$. This theoretical analysis provides an understanding about the driving mechanism of the enhanced modulation and is consistent with experimental measurements shown in Fig 3.

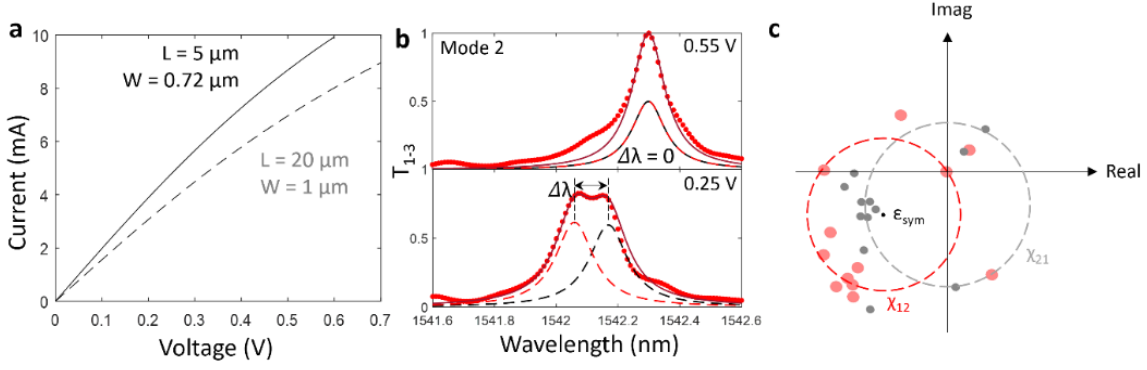

**Fig. S3 | Dynamic tuning of Hermiticity with local electro-optic phase shifter.** (a) The designed local heater current-voltage characteristics. (b) Transmission spectra of a mode evolve with the driving voltage for the inter-scatter phase. (c) The real part and imaginary part of $\chi_{12}$ and $\chi_{21}$ for the above transmission spectra.

### 4.3 Chiral cavity energy built-up

The total photon number in the microring resonator is computed by the sum of the intensities of two eigenmodes $U = |a|^2 = |a_+|^2 + |a_-|^2$ where $a_+ = \frac{a_{cw}+a_{ccw}}{\sqrt{2}}$ and $a_- = \frac{a_{cw}-a_{ccw}}{\sqrt{2}}$. Therefore, the total cavity energy at steady state is given by

$$U_{cw\_in} = \gamma_{c1} A_{cw,in}^2 \frac{\gamma_t^2/4 + |\chi_{12}^v|^2}{\left(\frac{\gamma_t^2}{4} + \chi_{12real}\chi_{21real} - \chi_{12imag}\chi_{21imag}\right)^2 + \left(\chi_{12real}\chi_{21imag} + \chi_{12imag}\chi_{21real}\right)^2}$$

(S. 4-6)

when only CW input is present, and by

$$U_{ccw_{in}} = \gamma_{c2} A_{ccw,in}^2 \frac{\gamma_t^2/4 + |\chi_{12}^v \chi_{21}^v|^2}{\left(\frac{\gamma_t^2}{4} + \chi_{12real}\chi_{21real} - \chi_{12imag}\chi_{21imag}\right)^2 + \left(\chi_{12real}\chi_{21imag} + \chi_{12imag}\chi_{21real}\right)^2}$$

(S. 4-7)

When only CCW input is present. There is a difference in the energy stored in the resonator for CW and CCW inputs due to the strong coupling and asymmetric coherent interference between CW and CCW modes. As a result, a dissimilar amount of nonlinear resonance shift is observed when transmission spectra for CW and CCW inputs are compared. This in turn leads to asymmetry in the transmission spectra. From (**S.4-3** and **S.4-4**), we can derive the cavity energy contrast of CW and CCW excitations as ($\gamma_{c1}=\gamma_{c2}$ in our device):

$$\frac{U_{cw\_in}}{U_{cw\_in}} = \frac{\gamma_t^2/4 + |\chi_{12}^v|^2}{\gamma_t^2/4 + |\chi_{21}^v|^2}$$

(S. 4-8)

Equation (S. 4-7) suggests that the manipulation of power asymmetry in reflection, represented by the system's chirality, allows for the augmentation of nonlinear effects. The nonlinear resonance shift (dominated by the thermal effect in silicon) is proportional to the cavity energy. The ratio between resonance shift and input power has a large contrast for CW and CCW excitations, which originated from the cavity energy difference and chirality of the ring (equation S. 4-5). Such asymmetric field enhancement and cavity energy built-up are experimentally verified through the nonreciprocities and chiral all-optical switching (Fig. 4).

The chiral energy built-up is experimentally characterized in Fig. 4. We characterized the chiral nonlinear response and power-dependent transmission (nonreciprocity between ports 1 and 2 as in Fig. 4d). The non-reciprocity intensity range ($NRIR = \frac{U_{cw\_in}}{U_{cw\_in}}$) remains unchanged with detuning (δ, the ratio between laser-resonance offset and half linewidth of the resonator) (Fig. S4a). NRIR only depends on the asymmetric inter-mode coupling, and increases with chirality (near EP), without additional loss for introducing asymmetric couplings as Fano nonlinear resonator (Fig. S4b). Near EP, the chiral energy built-up/nonlinear transmission shift has a similar form as the EP-enhanced tuning effect: $NRIR = 1 + \frac{|\chi_{12}|^2}{\gamma_t^2/4}$. An isolation depth of up to 4dB has been achieved so far (Fig. S4c).

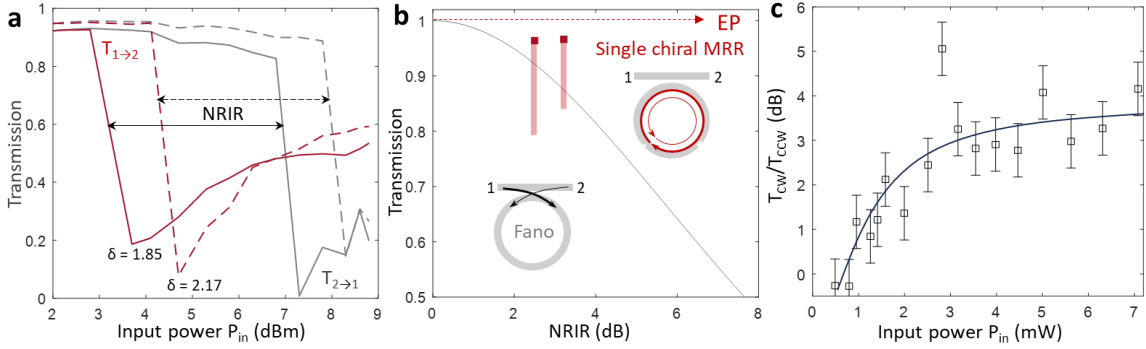

**Fig. S4 | High transmission nonreciprocal response near chiral EP. a.** Transmission versus input power from port 1 to 2 (red) and from port 2 to 1 (grey) at different detuning. **b.** Transmission versus NRIR for a single chiral (red dashed line) and Fano resonator (grey dashed line). Left inset: asymmetric coupling coefficients between feeding ports and resonator resulted in nonreciprocity in the Fano resonator, which resulted in the trade-off between transmission and NRIR. Right inset: asymmetric coupling between CW and CCW modes resulted in nonreciprocity in a chiral MRR. Lines and curves are theoretical predictions and squares are experimental data. **c.** Transmission contrast between CW and CCW excitations under the same input power. Squares are experimental data, and the curve is an eye guide.

**4.4 Comparison between nonlinear CMT and experimental data**

The nonlinear time domain CMT (**S.**4-1 ~ 4-2) derives the chiral energy built-up [S5]. The ensuing analysis is performed by Adams-Bashforth-Moulton method [S6].

$$\frac{da_{cw}}{dt} = -(j(\omega_L - \omega_0 + \Delta\omega_{nonlinear}) + \frac{1}{2\tau_t})a_{cw} - j\chi_{12}a_{ccw} - r_{c1}\sqrt{P_1} - r_{c2}\sqrt{P_3} \quad \textbf{(S. 4-9)}$$

$$\frac{da_{ccw}}{dt} = -\left(i(\omega_L - \omega_0 + \Delta\omega_{nonlinear}) + \frac{1}{2\tau_t}\right)a_{cw} - i\chi_{21}^v a_{cw} - r_{c1}\sqrt{P_2} - r_{c2}\sqrt{P_4} \quad \text{(S. 4-10)}$$

$$\frac{dN}{dt} = \frac{1}{2\hbar\omega_0 n_g^2}\frac{\beta_{si}c^2}{V_{FCA}^2}(U^2) - \gamma_{fc}N \quad \text{(S. 4-11)}$$

$$\frac{d\Delta T}{dt} = \frac{1}{\rho_{si}c_{p,si}V_{cavity}}P_{abs} - \gamma_{th}\Delta T \quad \text{(S. 4-12)}$$

where $N$ is the free-carrier density and $\Delta T$ is the cavity's temperature shift. The time-dependent resonant shift of the cavity is noted by $\Delta\omega_{nonlinear} = \Delta\omega_N - \Delta\omega_T$, where the free carrier dispersion is $\Delta\omega_N = \omega_0(\xi_e N + \xi_h N^{0.8})/n_{si}$. The thermally induced dispersion is $\Delta\omega_T = \omega_0 \Delta T (dn_{si}/dT)/n_{si}$. Kerr dispersion is negligibly small compared to the thermal and free-carrier mechanisms. The total loss rate dependent on input power is written by: $\gamma_t = (\gamma_{c1} + \gamma_{c2} + \gamma_{in}) + (\gamma_{lin} + \gamma_{TPA} + \gamma_{FCA})$, where the linear absorption $\gamma_{lin}$ for silicon is demonstrated to be small, $\gamma_{in}$ is the intrinsic loss which is related to the propagation loss, and $\gamma_{c1/2}$ is the coupling loss from the bus and drop waveguide. The free carrier absorption rate is given by $\gamma_{FCA} = c\sigma N(t)/n_g$. The field-averaged two-photon absorption rate is $\gamma_{TPA} = b_2 c^2/n_g^2/V_{TPA}|U(t)|^2$, where $b_2$ is the effective two-photon absorption coefficient. The total absorbed power is computed by $P_{abs} = (\gamma_{lin} + \gamma_{TPA}(U) + \gamma_{FCA}(U^2))U$.

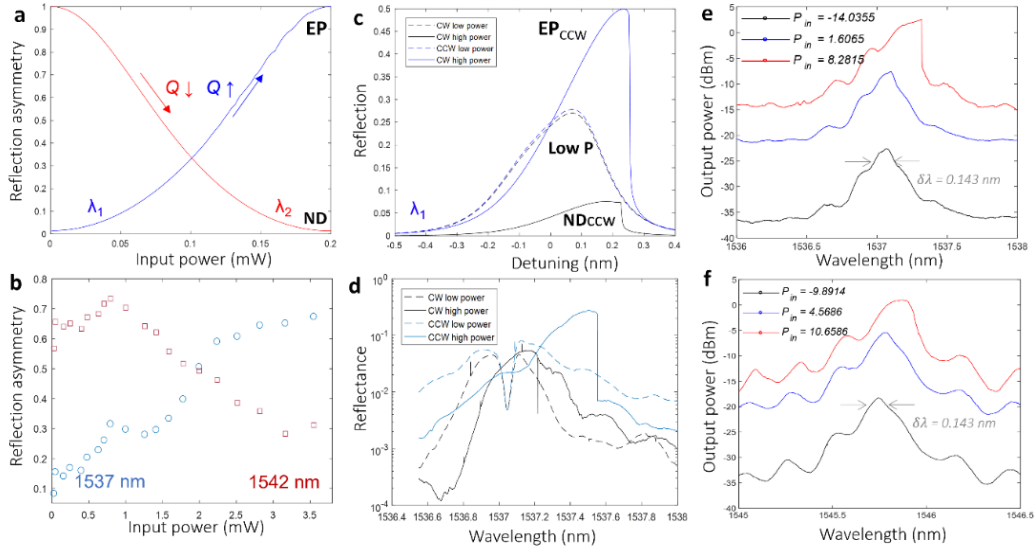

**Fig. S5 | Comparison between nonlinear CMT and experiment**. (a) Simulated and (b) measured optically tunable reflection asymmetry where reflection asymmetry = $(R_{2-3} - R_{1-4})/(R_{2-3} + R_{1-4})$ for two different resonant modes' cases where the reflection asymmetry increases with increasing power for $\lambda_1$ mode ($Q_{NH}\uparrow$), and it decreases for $\lambda_2$ mode ($Q_{NH}\downarrow$). (c) Simulated spectra of one resonance with CW mode evolved from DP to EP and CCW mode evolved from DP and ND with increased input power. Dashed curves: reflection spectra at low power. Solid curves: high power. (d) Experimentally measured spectra for the mode of resonance wavelength around 1537 nm in b. (e) EP and (f) non-EP transmission spectra from low optical power to high optical power with the same quality factor.

**TABLE S1** Estimated physical parameters from time-dependent coupled-mode theory-experimental matching, three-dimensional numerical field simulations, and measurement data. $\tau_{fc}$ is the effective free-carrier lifetime accounting for both recombination and diffusion.

| Parameter | Symbol | Si microring |
|---|---|---|
| TPA coefficient | $\beta_2$ ($10^{-11}$ m/W) | 0.84 |
| Kerr coefficient | $n_2$ (m$^2$/W) | $0.44 \times 10^{-17}$ |
| Thermo-optic coeff. | $dn/dT$ | $1.86 \times 10^{-4}$ |
| Specific heat | $c_v \rho$ (W/Km$^{-3}$) | $1.63 \times 10^6$ |
| Thermal relaxation time | $\tau_{th,c}$ (μs) | 1 |
| Thermal resistance | $R_{th}$ (K/mW) | 50 |
| FCA cross section | $\sigma$ ($10^{-22}$ m$^3$) | 14.5 |
| FCD parameter (electron) | $\zeta$ ($10^{-28}$ m$^3$) | 8.8 |
| FCD parameter (hole) | $\zeta$ ($10^{-28}$ m$^3$) | 4.6 |
| Carrier lifetime | $\tau_{fc}$ (ns) | 0.45 |
| Two photon absorption volume | $V_{TPA}$ ($10^{-18}$ m$^3$) | 35.15 [FDTD] |
| Free carrier absorption volume | $V_{FCA}$ ($10^{-18}$ m$^3$) | 31.7414 [FDTD] |
| Effective cavity mode volume | $V_{cavity}$ ($10^{-18}$ m$^3$) | 37.7594 [FDTD] |

**Supplementary Section 5: Nanofabrication and Characterizations**

**5.1 Design and fabrication of the photonic layer**

Fig. S6 illustrates the details of the fabrication and characterization of the photonic layer. Fig. S6a shows a dark field image of the chiral microring resonator array with add-drop filters (for obtaining the transmission and reflection in Fig. 3), where an inversely designed Y junction coupler was introduced to measure the CCW light injected reflection spectrums. SEM image of the Y junction shows the well-defined nanostructure in Fig. S6b. The inverse design method is performed based on the 'adjoint method' implemented in Python. And measured Y junction performance verifies the ideal 3dB loss as a 50:50 power splitter as shown in Fig. S6d. Those low-loss components are critical for nonlinear measurements, where the transmission is strongly dependent on the input power. The fiber-to-fiber loss of the single-step etched apodised grating coupler (without 3dB beam splitter) is ~10dB (Fig. S6c, d). The grating coupler loss per facet was estimated to be 5 dB near 1539 nm.

Fig. S6e-g illustrates the fabricated asymmetric Mie scatterers embedded in the micro-resonator. We prepared asymmetric Mie scatterers with varying dimensions from 200 nm to 100 nm (Fig. S6f,g). The smaller features around 100 nm show better performance. The design-fabrication variations are characterized to be around 5nm. The offset is affected by the adjacent topologies (Fig. S6g). The chirality measurement setup and nonlinear measurements are carried out using the setup illustrated in Fig. S7.

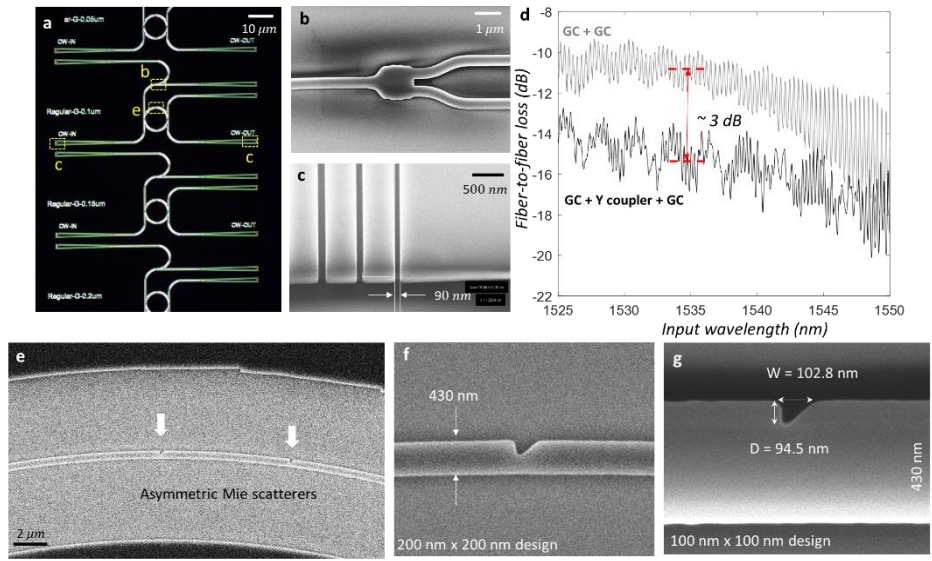

**Fig. S6 | Fabricated add-drop filter type ring resonator for obtaining the results in Fig. 3. a.** Dark field image of the fabricated photonic layer with add-drop filter around the chiral microring. The geometry of Mie-scatterer pair (embedded in the ring) achieving EP is illustrated in Fig. S1. **b.** SEM image of the inversely designed Y junction coupler, and **c.** apodised grating coupler, achieving 5dB insertion loss with a single step etch. **d.** Experimentally measured transmission spectrums of grating couplers (GC) with and without Y junction. **e.** The portion of the chiral microring with a pair of Mie scatterers. **f.** Zoom-in image of one design and **g** the other design we used for obtaining the results in Fig. 3.

**Fig. S7 | Testing setup for the chiral micro-resonator. a.** Set up schematics for obtaining the chiral transmission spectra. TLD: tunable laser diode. FPC: fiber polarization controller. PD: photodetector. O/C: switchable optical circulator. The chip-device coupling is unchanged during the measurement. The direction of optical excitation is controlled by the O/C. O/I: optical isolator. **b.** Device layout and testing ports.

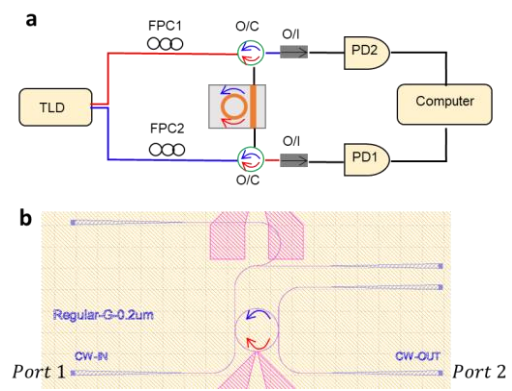

### 5.2 Precision alignment of nano-heater to nanophotonic structures

The fabrication flow chart for the nanophotonic layer and subsequent electronic layer is illustrated in Fig. S8a-b. The fabrication on the silicon layer defines the position and geometry of the microring and Mie scatterers (Fig. S8c), and the metal layer is precisely aligned with the photonic layer for achieving localized heating (Fig. S8d-f). Bi-layer resist (CSAR and LOR3A) was used for the better sidewall of the metal electrode than the traditional lift-off process. Electro-beam lithography alignment is a significant step to place the metal on the small area of the waveguide. Under the microscope, we place the Faraday cup to be the center of the optical microscope with the closely adjusted focal length. Then, we normalize the relative position for the Faraday cup and then shift the stage to see the chip in the middle of the microscope. Then we measure the four points of the chip edge in xy plane and calculate the chip position in x-y direction. These numbers are substituted in EBPG computer before starting the E-beam writing. SEM image of the fabricated microheater in Fig. S8d-f shows the nano heater can be aligned to the waveguide under the 700nm oxide cladding, with an offset of less than 5 nm. The fabricated microheater features a narrowest width of 50 nm with low edge roughness (Fig. S8f). The low edge roughness of the nano-scale metal layer is attributed to the bi-layer resist. We placed the chip in NMP solution for at least 1 hour at 80 degrees for the entire removal of the unexposed areas under E beam writing.

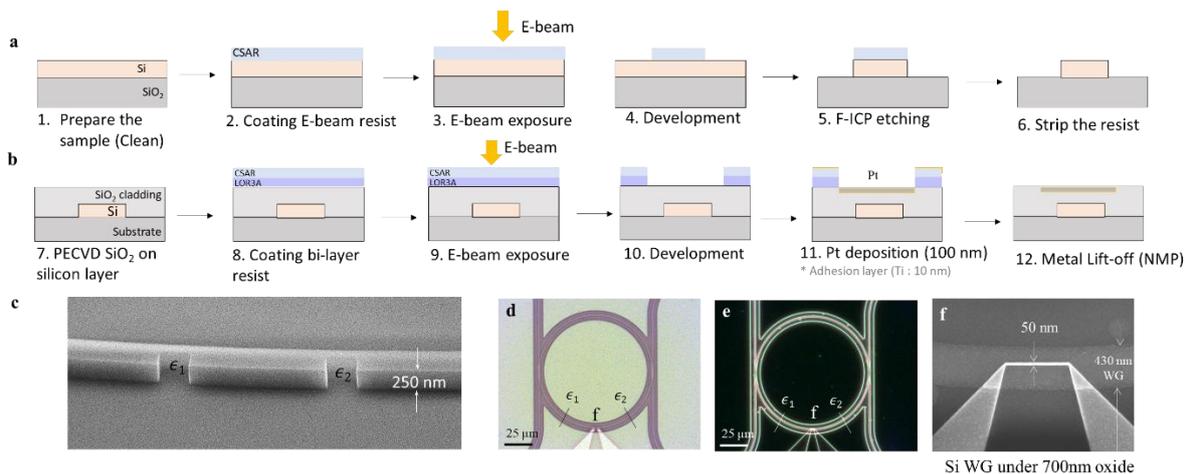

**Fig. S8 | Schematic of the active device fabrication process and SEM image of the fabricated narrowest micro-heater on top of the silicon ring resonator. a.** Fabrication process of the photonic layer with a single layer resist, and **b**, nano-scale electrodes aligned to the photonic layer, with double layer resist for achieving high precision and nanoscale heater. **c.** SEM image of an exemplary Mie scatterer pair embedded in the microresonator. **d.** Bright-field and **e** dark field image of the chiral microring with integrated heater. **f.** Achieved metal heater was 50 nm, and the heater design-fabrication offset is less than + 15 nm. *The alignment error between the heater and the middle of the waveguide (430nm wide) is less than 5 nm.*

The oxide cladding thickness is optimized by balancing the optical and heater designs. Through numerical full-field optical simulations (Fig. S9a-b), a minimal thickness of 700 nm oxide cladding is selected to prevent the waveguide mode coupling to the metal heater layer. Thinner oxide cladding also facilitates more efficient heating near the silicon waveguide. We characterize the heater designs with different widths. The heating efficiency of $\Delta T/P \sim 9.1$

K/mW is obtained through the current-voltage curve and temperature dependence of the Pt resistivity (Fig. S9c). Note that geometric engineering such as serpentine structures or introducing undercut can improve the heating efficiency, but those structures are too bulky for localized heating. From the thermal simulation (Fig. S9b), the tuning efficiency near the Si waveguide region is nearly half of the one on the top metal layer. The thermos-optic tuning efficiency is around 4.5 K/mW, which is more than two times the value reported in typical microring heaters (attributed to the nanoscale design) [S7-S8].

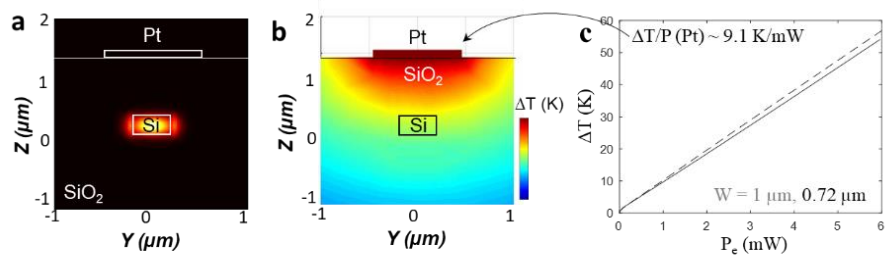

**Fig. S9 | Nanoscale local heater design and characterization**. **a.** Simulated optical field confided in the waveguide of the ring and **b,** Simulated temperature distribution from the top Pt heater on SOI substrate. **c.** Measured local temperature versus input power for the nano-heaters. The local temperature is derived from the measured current (I) and voltage (V) curve (Fig. S2a), with temperature dependence of the resistivity. The tuning efficiency on the metal layer is measured to be 9.1K/mW, and the tuning efficiency at the Si waveguide layer is around 4-5 K/mW.

**Supplementary Section 6: Nanomanufacturing chiral micro-resonator**

**6.1 Nanomanufacturing of tunable chiral micro-resonator**

With the advancement of technology nodes of the semiconductor photonic foundry, deep UV photolithography can support the nanomanufacturing of the chiral micro-resonator concept, with immersion lithography and fine-tuned etching steps. The critical dimension can reach 50 nm, as shown in Fig. S10a. The intrinsic quality factor of the microring resonator (with minimal surface roughness) decreased from $10^5$ to $10^4$ with the Mie scatterer (which is likely attributed to mode-splitting rather than additional radiation loss). Tuning such chiral microring towards EP requires additional electronic layers. Both photonic and electronic designs are adjusted and optimized complying with the foundry requirements. Two specially designed Mie scatterers (symmetric + asymmetric in Fig. S10a) with the same depth are utilized here, to ensure the same perturbation strength given the unknown geometric offset. The distance between two scatterers is approximately 30μm hosts the inter-scatterer phase tuner. We implemented structures of micro-heaters (Fig. S10b-c) and local *p-n* junction for non-thermal and localized phase tuning and modulation (Fig. 10d).

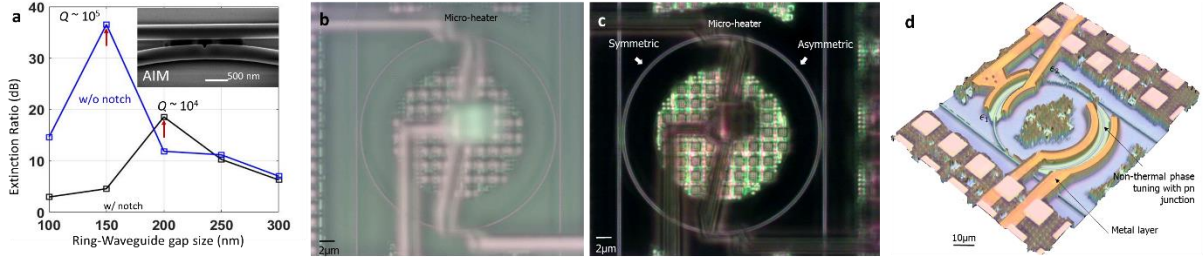

**Fig. S10 | Nanomanufactured Mie scatterer loss characterization, with the local phase tuning schemes of the micro-heater and *p-n* junction.** (a) The extinction ratio of foundry-manufactured notched MRRs versus ring-WG gap size (black), compared to the same design without notch (blue). Red arrows indicate the critical point. Inset: SEM image of the scatterer fabricated by 193nm deep-UV photolithography. Note that the foundry manufactured MRR has extremely low surface roughness. **b.** Bright field and **c.** dark field image of the chiral microring with a pair of matched Mie scatterers, and local heaters. **d.** Alternative design with inter-scatterer phase tuned by *p-n* junction. The refractive index change through carrier dispersion is highly localized.

For reducing the optical loss, the p-type doping concentration in the chiral microring modulator is $2\times10^7$ cm$^{-3}$, while the n-type doping concentration is $6\times10^7$ cm$^{-3}$. Heavy and heavily doped regions are utilized for achieving Ohmic contact. Due to the short distance of the *p-n* junction in the chiral microring modulator, the change in the refractive effective index is on the order of $10^{-4}$. Additionally, the loss in the waveguide resulting from this structure is approximately $10^{-2}$ dB. These values feature the sensitivity and efficiency of chiral modulator in terms of its refractive index modulation and the accompanying loss in the waveguide. The estimated total quality factor, determined through a single Lorentzian fit process, is approximately $\sim10^4$. In Fig. S11a, the electro-optical tuning measurements are used to explain the chiral modulation behavior. A specific mode within the brown dashed box from Fig. S11b is selected for analysis. Fig. S11c provides the time domain modulation results at 1555.85nm. With the same electrical drive (black curves), CW (blue) and CCW (red) excitation exhibit distinguished responses, demonstrating directional electro-optic modulation.

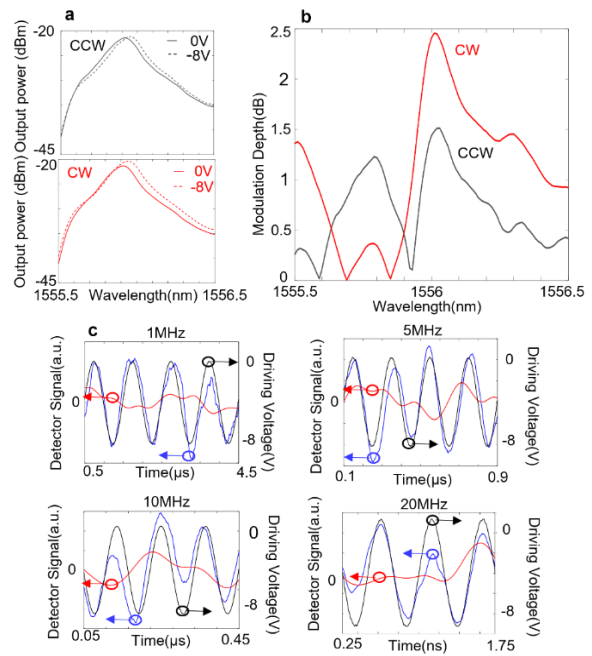

**Fig. S11 | Electro-optical tuning analysis and temporal modulation results.** (a) DC performance of the CW and CCW reversed

biased ring modulator. (b) Modulation depth (OMA) versus wavelength. (c) Time domain experimental results in reversed bias with 1MHz, 5MHz, 10MHz, and 20MHz. Driving signal: black solid line; CW: red solid line; CCW: blue solid line.

**6.2 Carrier plasma-induced local phase shift and associated mode splitting**

In this study, we utilized double Lorentzian formulas to fit the forward bias experimental data for both CW and CCW excitations, as depicted in Fig. S12. Through the fitting process, we successfully extracted crucial parameters, including the total quality factor ($Q_t$), intrinsic quality factor ($Q_{in}$), and coupling quality factor ($Q_c$) from the obtained results. Additionally, the mechanism of the EP ring resonator is explained using $\frac{1}{Q_t} = \frac{1}{Q_c} + \frac{1}{Q_{in}} + \frac{1}{Q_{NH}}$. For both directions, $Q_c$ and $Q_{in}$ remain unchanged. External bias modifies $Q_{in}$ through carrier absorption and dispersion effect, and $Q_{NH}$ through inter-mode coupling. $Q_{in}$ includes radiation loss and material loss. Further p-n junction engineering can reduce carrier absorption associated $Q_{in}$ reduction, for improved performance with high-speed operation.

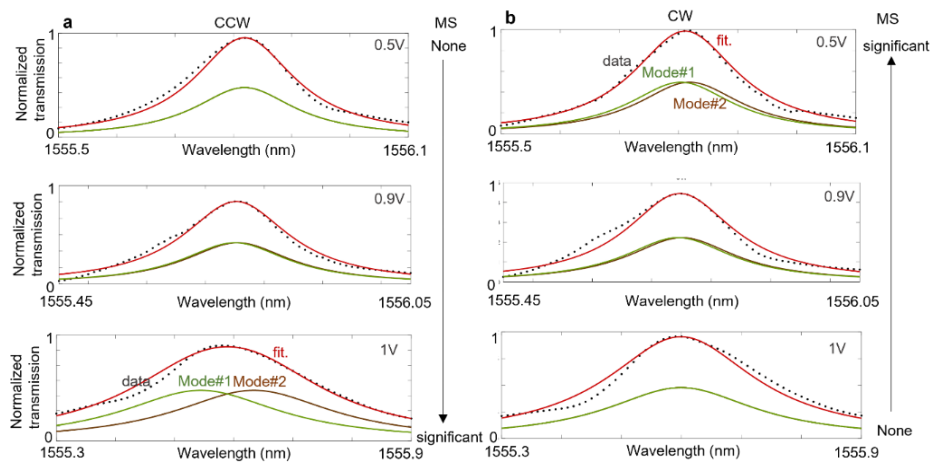

**Fig. S12 | Chiral response of mode-splitting on transmission ports.** Transmission spectra were obtained for (a) CCW and (b) CW excitations. The dotted line is experimental data. Brown and green solid curves represent the modes inside the resonance. Red solid curves are the convolution of the two Lorentzian components. MS: mode-splitting.